\pdfoutput=1

\documentclass[11pt]{article}

\usepackage[]{acl}

\usepackage{times}
\usepackage{latexsym}
\usepackage{graphicx}
\usepackage{amssymb}
\usepackage{amsmath}
\usepackage{array,multirow}
\usepackage{subcaption}
\usepackage{tabularx}
\usepackage{pbox}
\graphicspath{ {./images/} }
\usepackage[T1]{fontenc}

\long\def\comment#1{}
\long\def\comments#1{}

\usepackage{microtype}
\newcommand*{\Resize}[2]{\resizebox{#1}{!}{$#2$}}%
\title{
Compact Token Representations with Contextual Quantization for Efficient Document Re-ranking
}

\author{Yingrui Yang, Yifan Qiao, Tao Yang\\
  Department  of Computer Science, University of California at Santa Barbara, USA \\
  \texttt{\{yingruiyang, yifanqiao, tyang\}@cs.ucsb.edu} }

\begin{document}


\maketitle

\begin{abstract}
Transformer based re-ranking models can achieve  high search relevance through 
context-aware soft matching of query tokens with document tokens. 
To alleviate runtime complexity of such inference, previous work has adopted a late interaction architecture
with pre-computed contextual token representations at the cost of a large online storage.
This paper proposes contextual quantization of token embeddings by 
decoupling document-specific and document-independent ranking contributions
during codebook-based compression. This allows effective online decompression and embedding composition for better search relevance.
This paper presents  an evaluation of 
the above compact token representation model in terms of relevance and space efficiency.

\end{abstract}

\comments{
Transformer based models have achieved state-of-the-art performance on question answering and ranking tasks. Specifically, dual encoders designs that separately encode a pair of texts for ranking is widely adopted. However, in order to alleviate the expensive computation of BERT for texts in online system, pre-computation of contextual token representations are required. As a token in different context has different contextual representations, the storage requirement for storing such representation is tremendous. In this work, we propose a neural network based approach to decouple the word sense and word static representation and learns a compositional code embedding for word senses. We demonstrate the compression effectiveness on a state-of-the-art dual encoder ranking model. Through this approach, we are able to reduce the storage of contextual token representation by XX times with minimum degradation in model effectiveness.
}

\section{Introduction}
Modern search engines for text documents
typically employ multi-stage ranking. 
The first retrieval stage extracts top candidate
documents matching a query from a large search index with a simple  ranking method. 
The second stage or a later stage uses a more  complex machine learning algorithm to 
re-rank top results thoroughly.
Recently neural re-ranking techniques from transformer-based architectures
have achieved impressive relevance scores 
for top $k$ document re-ranking, such as ~\citet{ MacAvaney2019SIGIR-cedr}. 
However, using a transformer-based  model to rank or re-rank 
is extremely expensive during the online  inference~\cite{Lin2020PretrainedTF}. Various
efforts have been made to reduce its computational complexity (e.g. ~\citet{Gao2020distill}).

A noticeable success in time efficiency improvement is accomplished  in 
ColBERT~\cite{colbert} which conducts late interaction of query terms
and document terms during runtime inference so that token embeddings for documents 
can be pre-computed.
Using ColBERT re-ranking after a sparse retrieval model called DeepImpact~\citep{Mallia2021deepimpact} 
can  further  enhance relevance. 
Similarly BECR~\citep{Yang2021WSDM-BECR}, CEDR-KNRM~\citep{MacAvaney2019SIGIR-cedr}, and PreTTR~\citep{MacAvaney2020SIGIR-prettr} have 
also adopted  the late interaction architecture in their efficient transformer based re-ranking schemes.

While the above work  delivers good search relevance with late interaction, 
their improvement in time efficiency  has come at the cost of a large storage space in hosting token-based precomputed document 
embeddings. For example, for 
the MS MARCO document corpus, the footprint of embedding vectors in 
ColBERT takes up to 1.6TB and hosting them in a disk incurs substantial time
cost when many embeddings are fetched for re-ranking.
\comments{
ClueWeb\footnote{https://lemurproject.org/clueweb09/} with 50 million documents, even with LSH compression for space reduction, the cost storing contextual term-based document embeddings in BECR exceeds 7 terabytes. 
For the same dataset, ColBERT under the smallest recommended configuration would cost 4.1 terabytes.
}
It is highly desirable to reduce embedding footprints and host them in memory as much as possible for fast and high-throughput
access and for I/O latency and contention avoidance, especially when an online re-ranking server is required to efficiently process  many queries simultaneously.

The {\bf contribution} of this paper is to propose a compact representation for contextual token
embeddings of documents called Contextual Quantization (CQ). 
Specifically, we adopt codebook-based quantization to 
compress embeddings 
while explicitly  decoupling the ranking contributions of  document specific   
and document-independent information in contextual embeddings.
These ranking contributions are recovered with weighted composition after quantization decoding  during online inference.
Our CQ scheme includes a neural network model that jointly learns context-aware decomposition and quantization with 
an objective to preserve correct ranking scores and order margins. 
Our evaluation shows that CQ can effectively reduce the storage space of contextual representation by about 14 times 
for the tested  datasets
with insignificant online embedding recovery overhead and a small relevance degradation for re-ranking  passages or documents.

\section{Problem Definition and Related Work}
\label{sect:background}

The problem of neural text document re-ranking is defined as follows.
Given a query with multiple terms and a set of candidate documents,
rank these documents mainly based on their embeddings and query-document similarity. 
With a BERT-based re-ranking algorithm, typically a term is represented by a token,
and thus in this paper,  word 	``term'' is used interchangeably with ``token''.
This paper is focused on minimizing the space cost  of token embeddings for fast online re-ranking inference.


\textbf{Deep contextual re-ranking models}.
Neural re-ranking has pursued representation-based or interaction-based algorithms~\cite{drmm, convknrm, Xiong2017SIGIR-knrm}.
Embedding interaction based on query and document terms shows an advantage in these studies.
The transformer architecture based on BERT ~\cite{Devlin2019BERT} has been adopted
to re-ranking tasks by using  BERT's
[CLS] token representation to summarize query and document interactions~\cite{Nogueira2019PassageRW,birch,bertmaxp, monobert, parade}. 
Recently BERT is  integrated in late term interaction~\cite{MacAvaney2019SIGIR-cedr,2020ECAI-tk,tkl,2021SIGIR-ck-Mitra} which delivers
strong relevance scores for re-ranking. 

\textbf{Efficiency optimization for  transformer-based re-ranking.}
Several approaches have been proposed to reduce the time complexity of transformer-based ranking.  
For example, architecture simplification~\cite{2020ECAI-tk,2021SIGIR-ck-Mitra}, late interaction with
precomputed token embeddings~\cite{MacAvaney2020SIGIR-prettr}, early exiting~\cite{Xin2020SUSTAINLP},
and  model distillation~\cite{Gao2020distill, Hofsttter2020marginMSE, Chen2020-SimplifiedTK}.
We will focus on  the compression of  token  representation following the late-interaction work of
ColBERT~\citep{colbert} and BECR~\citep{Yang2021WSDM-BECR} as they deliver fairly competitive relevance scores for several well-known ad-hoc TREC datasets.
These late-interaction approaches follow a  dual-encoder design that separately encodes the two
sets of texts, studied in various NLP
tasks~\citep{Zhan2020LearningTR,Chen2020DiPairFA,Reimers2019SentenceBERTSE,Karpukhin2020DensePR,dcbert}.

Several previous re-ranking model attempted to reduce the space need for contextual token embeddings.
ColBERT has considered an option of using a smaller dimension per vector and limiting 2 bytes per number as a scalar quantization. 
BECR~\cite{Yang2021WSDM-BECR} uses LSH for hashing-based contextual embedding compression~\cite{JiWWW2019}.
PreTTR~\cite{MacAvaney2020SIGIR-prettr} uses a single layer encoder model to reduce the dimensionality of  each token embedding. 
Following PreTTR, a contemporaneous work called SDR  in~\citet{2021-SDR-Cohen}  
considers an autoencoder to reduce the dimension of representations, followed by an off-the-shelf scalar quantizer. 
For the autoencoder, it combines static BERT embeddings with contextual embeddings. 
Inspired by this study, our work decomposes contextual embeddings to decouple ranking contributions during vector quantization. 
Unlike SDR, CQ jointly learns the codebooks and decomposition for the document-independent and dependent components guided by a ranking loss.

\textbf{Vector quantization.}
Vector quantization with codebooks 
was  developed  for data compression to assist  approximate  nearest neighbor search, for example, 
product quantizer (PQ) from \citet{Jgou2011PQ}, optimized product quantizer (OPQ) from \citet{Ge2013OPQ};
residual additive quantizer(RQ) from~\citet{Ai2015RQ} and local search additive quantizer (LSQ) 
from~\citet{Martinez2018ECCV-LSQ}.
Recently such a technique has been used for compressing static word embeddings~\cite{Shu2018ICLR}
and document representation vectors in a dense retrieval scheme called JPQ~\cite{2021CIKM-JPQ-Zhan}. 
None of the previous work has worked on quantization of contextual 
token vectors for the re-ranking task, and that is the focus of this paper.


\section{Contextual Quantization}
\comments{
\begin{figure*}[h!]
    \centering
    \includegraphics[scale=0.5]{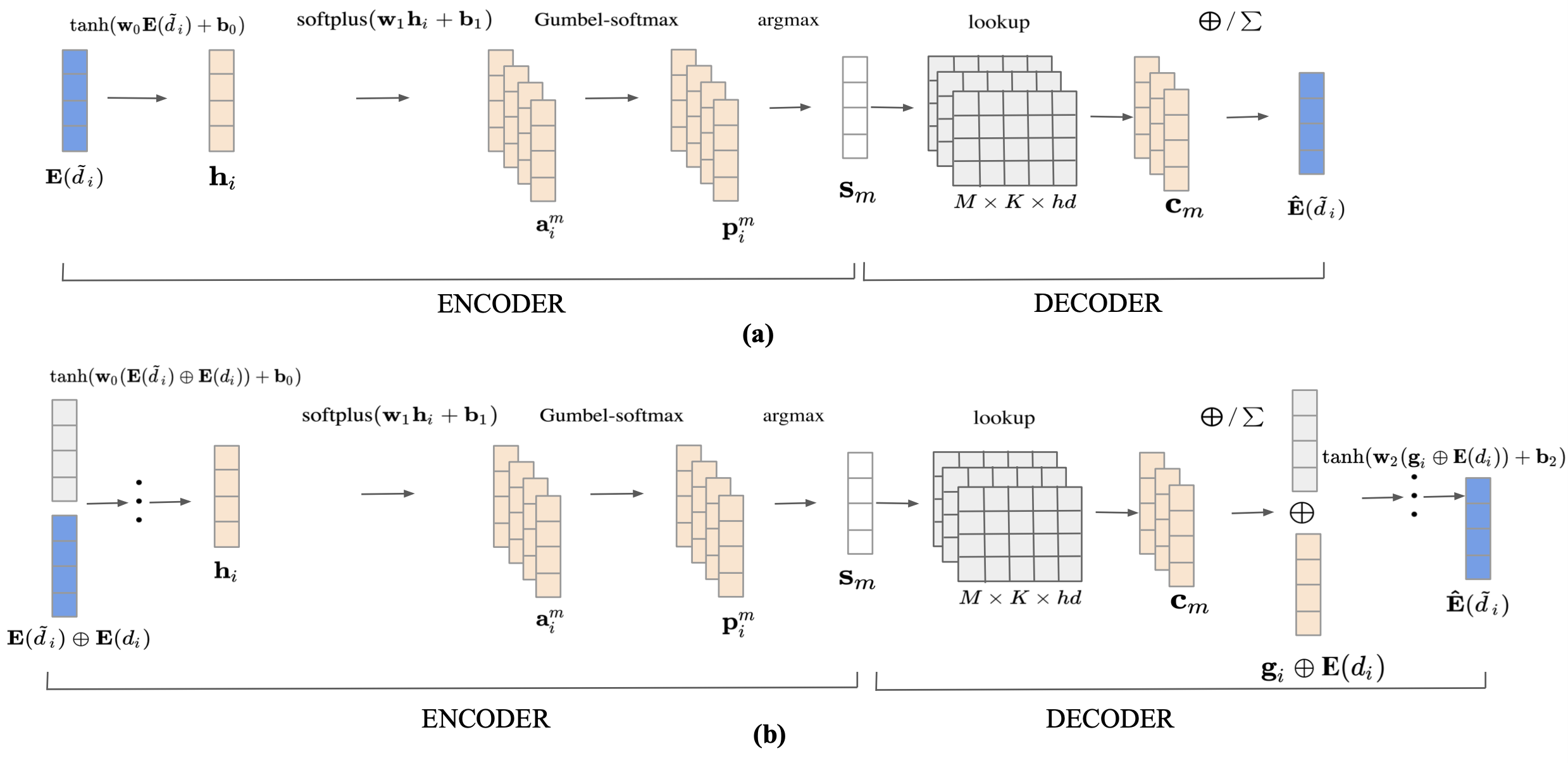}
    \caption{(a) Vector quantizer without contextual decomposition; (b) Contextual quantization.}
    \label{fig:model}

\end{figure*}
}

\begin{figure*}[h!]
    \centering
    \includegraphics[scale=0.35]{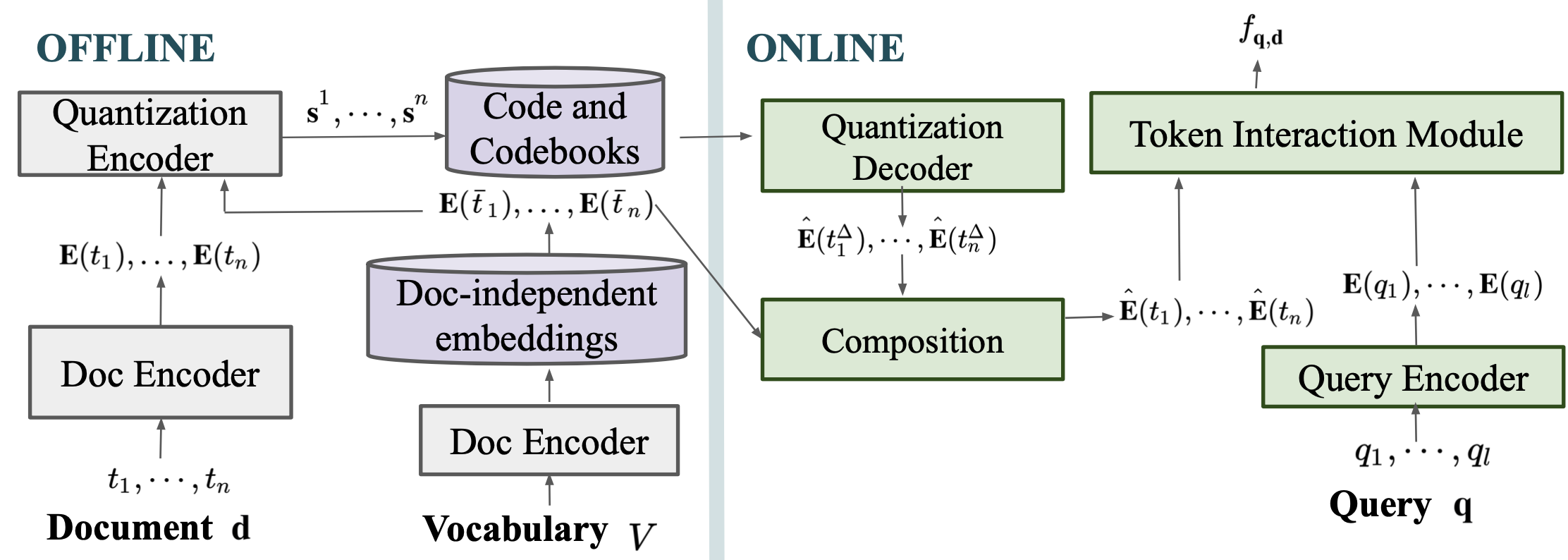}
    \caption{Offline processing and online ranking with contextual quantization}
    \label{fig:ranking}
\end{figure*}

Applying  vector quantization  naively to token embedding compression  does not ensure the 
ranking effectiveness 
because  a quantizer-based compression  is not lossless, and  critical ranking signals could be lost during data transformation.
To achieve a high compression ratio while maintaining the competitiveness in  relevance,
we consider the  ranking contribution of a contextual token embedding for soft matching containing two components:
1) document specific component derived from the self attention among context in a document,
2) document-independent and corpus-specific
 component generated by the transformer model. 
Since  for a reasonable sized document set, the second component is invariant to documents, its storage space is 
negligible compared to the first component. Thus the second part
 does not need compression. We focus on compressing the first component using codebooks.
This decomposition strategy can reduce the relevance loss due to compression approximation, which allows a more aggressive compression ratio.
Our integrated vector quantizer with contextual decomposition contains a ranking-oriented scheme with an 
encoder and decoder network for jointly learning codebooks and composition weights. Thus, the online composition
of decompressed document-dependent information with document-independent information can  retain a good relevance.

\subsection{Vector Quantization and Contextual Decomposition }
\label{sec:method-NC}

A vector quantizer consists of two steps as discussed in~\citet{Shu2018ICLR}. 
In the compression step, it encodes a real-valued vector (such as a token embedding vector in our case) 
into a short code using a neural encoder. The short code is a list of reference indices to the codewords in codebooks. 
During the decompression step, a neural decoder is employed  to reconstruct the original vector from the code and codebooks. 

The quantizer learns a set of $M$ codebooks $\{\mathcal{C}^1,\mathcal{C}^2,\cdots,
\mathcal{C}^M\}$ and  each codebook contains $K$ codewords ($\mathcal{C}^m = \{\textbf{c}^m_1$, $\textbf{c}^m_2$, $\cdots$, $\textbf{c}^m_K\}$) of dimension $h$. 
Then for any D-dimensional real valued vector $\textbf{x} \in \mathbb{R}^D$, the encoder 
compresses $\textbf{x}$ into an $M$ dimensional code vector $\textbf{s}$. Each entry
of code $\textbf{s}$ is an integer $j$, denoting the $j$-th codeword in codebook $\mathcal{C}^m$.
After locating all $M$ codewords as $[\textbf{c}^1, \cdots, \textbf{c}^M]$,
the original vector can be recovered with two options.
For a product quantizer
, the dimension of codeword  is $h = D / M$, and   
the decompressed vector is $\hat{\textbf{x}} = \textbf{c}^1 \circ  \textbf{c}^2 \cdots  \circ  \textbf{c}^M$ where symbol
$\circ $ denotes vector concatenation.
For an additive quantizer
the decompressed vector is $\hat{\textbf{x}} = \sum_{j=1}^{M} \textbf{c}^j$. 

\label{sec:method-WSD}

{\bf Codebook-based contextual quantization.}
Now we describe how codebook-based compression  is used in our contextual quantization.
Given a token $t$, we consider its contextual embedding vector $\textbf{E}(t)$  as  a weighted combination of two components: 
$\textbf{E}(t^{\Delta})$ and  $\textbf{E}(\bar{t})$.
$\textbf{E}(t^{\Delta})$ captures the document-dependent component, and
$\textbf{E}(\bar{t})$ captures the document-independent component discussed earlier.
For a transformer model such as BERT, 
$\textbf{E}(t)$ is the token output from the last encoder layer, and 
we obtain $\textbf{E}(\bar{t})$ 
by feeding $\text{[CLS]} \circ t \circ \text{[SEP]}$ into BERT model and taking last layer's output for $t$.

During offline data compression, we do not explicitly derive $\textbf{E}(t^{\Delta})$ 
as we only need to store the compressed format of such a value, represented as a code.
Let $\hat{\textbf{E}}(  t^{\Delta} )$ 
be the recovered vector with codebook-based 
decompression, as a close approximation of $\textbf{E}(t^{\Delta})$.
Let $\hat{\textbf{E}}(t)$ be the final composed embedding  used for online ranking with late-interaction.
Then
$\hat{\textbf{E}}(t)
= g( \hat{\textbf{E}}(t^{\Delta}),  \textbf{E}(\bar{t}) )$
where $g(.)$ is a simple feed-forward network to combine two ranking contribution components. 

{\bf The encoder/decoder  neural architecture  for contextual quantization.}
We denote a token in a document $\textbf{d}$ as $t$. 
The input to the quantization encoder is $\textbf{E}(t) \circ \textbf{E}(\bar{t})$. 
The output of the quantization encoder is the code vector $\textbf{s}$ of dimension $M$. 
Let code $\textbf{s}$ be $(s_1, \cdots, s_m, \cdots,  s_M)$ and each entry $s_m$ will be computed below in Eq.~\ref{eq:encoder_code}.
This computation uses the hidden layer $\textbf{h}$ defined as:
\begin{align}
    \textbf{h} = & \tanh(\textbf{w}_0 (\textbf{E}(t) \circ \textbf{E}(\bar{t})) + \textbf{b}_0) \label{eq:encoder_hidden}.
\end{align}
The dimension of $\textbf{h}$ is fixed as $1 \times MK/2$. 
The hidden layer $\textbf{a}$ is computed by a feed forward layer with a 
softplus activation (Eq. \ref{eq:encoder_hidden2}) with an output dimension of $M \times K$ after reshaping,
Let $\textbf{a}^m $ be the $m$-th row of this output.
\begin{align}
    \textbf{a}^m 
= & \text{softplus} (\textbf{w}_1^m \textbf{h} + \textbf{b}_1^m)\label{eq:encoder_hidden2}.
\end{align}
To derive a discrete code entry for $s_m$, following the previous work~\citep{Shu2018ICLR}, 
we apply the Gumbel-softmax trick~\citep{Maddison2017Gumbel,Jang2017Gumbel} 
as shown in Eq. \ref{eq:encoder_gumbel}, where the temperature $\tau$ is 
fixed at 1 and $\epsilon_k$ is a noise term sampled from the Gumbel 
distribution 
$- \log ( -\log 
(\text{Uniform}[0,1]))$. Here $\textbf{p}^m$ is a vector with dimension $K$. $(\textbf{p}^m)_j$ is the $j$-th entry of 
the vector. Similarly, $(\textbf{a}^m)_j$ is the $j$-th entry of $\textbf{a}^m$. 
\begin{align}
    (\textbf{p}^m)_j = &  \frac{\exp(\log ((\textbf{a}^m)_j + \epsilon_j)/\tau)}{\sum_{j' = 1}^K \exp(\log ((\textbf{a}^m)_{j'} + \epsilon_{j'})/\tau)}.\label{eq:encoder_gumbel}\\
    s_{m} & = \arg\max_{1\le j \le K}  (\textbf{p}^m)_j.\label{eq:encoder_code}
\end{align}
  
In the decompression stage,  the input to the quantization decoder is the code $\textbf{s}$, and this decoder accesses 
$M$ codebooks $\{\mathcal{C}^1,\mathcal{C}^2, \cdots, \mathcal{C}^M\}$ as  $M$ parameter matrices of size $K\times h$ which will be learned. 
For each $m$-entry of code $\textbf{s}$, $s_m$ value is the index of row vector in $\mathcal{C}^m$ to be used as its corresponding codeword.
Once all codewords $\textbf{c}^1$ to $\textbf{c}^M$ are fetched, we recover the approximate vector 
$\hat{\textbf{E}}(t^{\Delta})$ 
as $\sum_{j=1}^{M} \textbf{c}^j$ 
for additive quantization or $\textbf{c}^1 \circ \textbf{c}^2 \cdots  \circ \textbf{c}^M$  for product quantization. 

Next, we perform a composition with a one-layer or two-layer feed-forward network
to derive the  contextual embedding 
as $\hat{\textbf{E}}(t)= g(\hat{\textbf{E}}(t^{\Delta}, \textbf{E}(\bar{t}))$.
With one feed-forward layer, 
\comments{ 
concatenating $\hat{\textbf{E}}(t^{\Delta} )$ 
embedding $\textbf{E}(\bar{t})$ and go through one or two feed forward layers to 
get the approximated contextual embedding $\hat{\textbf{E}}(t)$ (Equation \ref{eq:composition}),
}
\begin{equation}
    \hat{\textbf{E}}(t) = \tanh(\textbf{w}_2 (
\hat{\textbf{E}}(t^{\Delta} )
 \circ\textbf{E}( \bar{t})) + \textbf{b}_2).\label{eq:composition}
\end{equation}

\comments{
\begin{figure}[h!]
    \centering
    \includegraphics[scale=0.32]{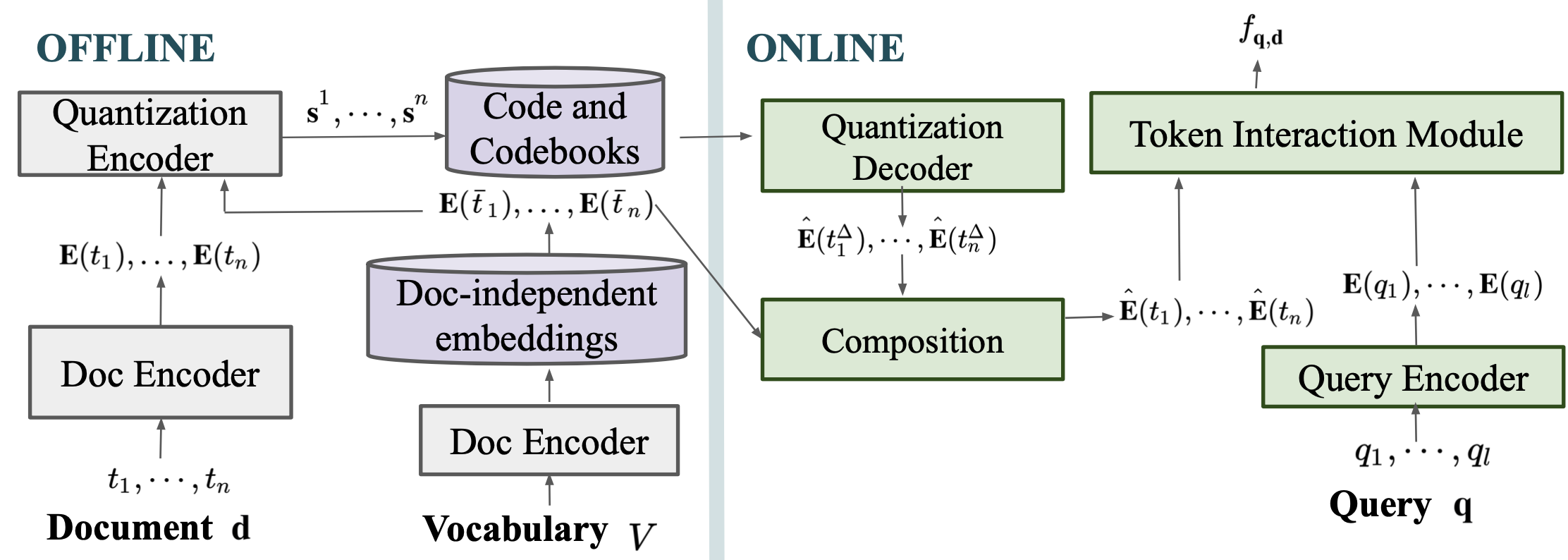}
    \caption{Offline processing and online ranking with contextual quantization}
    \label{fig:ranking}
\end{figure}
}

The above encoder and decoder for quantization
have parameter $\textbf{w}_0, \textbf{b}_0,\textbf{w}_1, \textbf{b}_1, \textbf{w}_2, \textbf{b}_2$, and
$\{\mathcal{C}^1,\mathcal{C}^2,\cdots, \mathcal{C}^M\}$. 
These parameters are learned through training. 
Once these parameters are learned, the quantization model is fixed and the code for any new token embedding
can be computed using Eq.~\ref{eq:encoder_code} in offline processing. 

Figure~\ref{fig:ranking} depicts the flow  of offline learning 
and the online inference with context quantization.
Given a query with $l$ tokens $\{q_1, q_2,..q_l\}$, and a 
documents with $n$ tokens $\{t_1, t_2,..t_n\}$, 
The query token embeddings  encoded with  a transformer based model (e.g. BERT)
are  denoted as $\textbf{E}(q_1), \cdots, \textbf{E}(q_l)$.
The  embeddings for document tokens through codebook base decompression are
$\hat{\textbf{E}}(t_1),
\cdots
\hat{\textbf{E}}(t_n)$.
The online inference then uses the interaction of query tokens and document tokens
defined in a re-ranking algorithm such as ColBERT
to derive a ranking score (denoted as $f_{\textbf{q},\textbf{d}}$).

The purpose of injecting $\textbf{E}( \bar{t})$  in Eq.~\ref{eq:encoder_hidden}
is to decouple the document-independent ranking contribution from  contextual embedding
$\hat{\textbf{E}}(t^{\Delta})$ so that this quantization encoder model will be learned to implicitly extract and 
compress the document-dependent ranking contribution.


Table~\ref{tab:example} gives an example with several token codes  produced by CQ for
 different sentences representing different contexts,
and illustrates context awareness of CQ's encoding with a small codebook dimension (M=K=4).
For example, 1 in code [4, 4, 3, 1] means the 4-th dimension uses the first codeword of the corresponding codebook.  
Training of CQ uses the MS MARCO passage dataset discussed in Section~\ref{sec:eval}
 and these sentences are not from this dataset.
Our observation from this example is described as follows.
First,  
in general token codes in the same sentences are closer to each other,
and token codes  in 
different sentences, even with the same word ``bank'',  are far away with a visible Hamming distance.
Thus CQ coding allows  a context-based separation among tokens residing in different contexts.
Second, by looking at boldfaced tokens at  each sentence, their distance in terms of  contextual semantics and proximity  is
reflected  to some degree in their CQ codes. 
For instance, a small Hamming code distance of three words ``actor'', ``poet'' and ``writer'' resembles  their semantic and positional  closeness. 
A larger code distance of two  ``bank''s in the 3$^{rd}$ and 4$^{th}$ sentences  relates with  their word sense and positional difference. 

 \begin{table*}[ht]
    \centering
    \resizebox{1.7\columnwidth}{!}{

    \begin{tabular}{l |lll}
        \hline
     \textbf{Context}  & \multicolumn{3}{c}{\textbf{Token codes} } \\
     \hline\hline
     William Shakespeare was widely regarded as the world's greatest     & writer & actor & poet  \\
      \textbf{actor}, \textbf{poet}, \textbf{writer} and dramatist.  & 
[4,4,3,1]& [4,4,3,1]& [1,4,3,1]\\
        \hline
     I would like to have either a cup of \textbf{coffee} or a good \textbf{fiction} & coffee & fiction & \\
     to kill time. &
 [3,3,3,4] & [3,1,3,4] & \\
        \hline
    She sat on the river \textbf{bank} across from a series of wide, & 1$^{st}$ bank & 2$^{nd}$ bank & \\
    large steps leading up a hill to the \textbf{bank} of America building. &
 [3,1,4,2] & [4,1,3,1] & \\
    \hline
    Some language techniques can recognize  word senses in phrases &1$^{st}$ bank &2$^{nd}$ bank& \\
    such as a river \textbf{bank} and a \textbf{bank} building. & [4,3,2,2] & [3,1,1,4] & \\
    \hline
    If you \textbf{get} a cold, you should drink a lot of water and \textbf{get} some rest. & 1$^{st}$ get & 2$^{nd}$ get& \\
    				& [2,2,4,2] & [2,1,2,4] & \\
        \hline
    \end{tabular}
    }
\caption{Example context-aware token codes produced by CQ using M=K=4  for the illustration purpose. }
\label{tab:example}
\end{table*}

\comments{
\begin{tabular}{l |llll}
    \hline
 \textbf{Context}  & \multicolumn{4}{c}{\textbf{Token} } \\
 & queen & frame & guitar  & [PAD]\\
 \hline\hline
 A careful application of data  preparation \underline{\hspace{0.5cm}} is required  in order to  avoid data &&&&\\
  leakage, and this varies depending on the model  evaluation scheme used. & [2,0,0,1]& [2,2,0,1]& [2,1,0,1] & [2,1,0,1]\\
    \hline
William Shakespeare was an English  playwright, poet, and actor, widely regarded   &&\\  
 as the greatest \underline{\hspace{0.5cm}}  in the English language and the world's greatest dramatist.& [0,0,0,1] & [0,0,3,1] & [0,0,3,1] & [0,0,3,1] \\ 
    \hline
\end{tabular}
}

\textbf{Training loss for parameter  learning}.
\label{sec:method-training}
We have explored three training loss functions. 
The first option is to follow a general quantizer~\citep{Shu2018ICLR} using
the  mean squared error (MSE) between the reconstructed and original embedding vectors of all token $t_i$. 
Namely $\mathcal{L}_{MSE} = \sum \|\textbf{E}(t_i)-\hat{\textbf{E}}(t_i)\|_2^2$.

The second option is the  pairwise cross-entropy loss based on rank orders.
 After warming up with the MSE loss, we further train the quantizer using 
$ \mathcal{L}_{PairwiseCE} =  \sum (-\sum_{j=\textbf{d}^+, \textbf{d}^-}P_{j} \log P_{j})$
where $\textbf{d}^+$ and $\textbf{d}^-$ are  positive and negative documents for query $q$.

We adopt the third option which
borrows the idea of MarginMSE loss from ~\citet{Hofsttter2020marginMSE} proposed for BERT-based ranking model distillation. 
In MarginMSE, a student model is trained to mimic the teacher model in terms of both ranking scores as well as the document relative order margins. 
In our case, the teacher model is the ranking model without quantization and the student model is the ranking model with quantization. It is defined as 
$ \Resize{7cm}{\mathcal{L}_{MarginMSE} = \sum ( (f_{\textbf{q},\textbf{d}^+}- f_{\textbf{q},\textbf{d}^-}) - (\hat{f}_{\textbf{q},\textbf{d}^+}- \hat{f}_{\textbf{q},\textbf{d}^-}) )^2 } $,
 where 
$f_{\textbf{q},\textbf{d}}$ 
and $\hat{f}_{\textbf{q},\textbf{d}}$ denote the ranking score with and without  quantization, respectively.
The above loss function distills the ColBERT ranking characteristics into the CQ model for better preservation of ranking effectiveness.

\subsection{Related Online Space and Time Cost}
\label{sec:method-storage}
\textbf{Online space for document embeddings.}
The storage cost of the precomputed document embeddings in a late-interaction re-ranking algorithm 
is dominating its online space need.
To recover token-based document embeddings, an online  server with contextual quantization 
stores three parts: codebooks,  the short codes of tokens in each document,  and the document-independent embeddings.

Given  a document  collection of $Z$ documents of length $n$ tokens on average, 
let $V$ be the number of the distinct tokens. For
$M$ codebooks with $M* K$ codewords of dimension $h$, 
we store each entry of a codeword with a 4-byte floating point number. 
Thus the space cost of codebooks is $M * K * h * 4$ bytes, and
the space  for document-independent embeddings of dimension $D$ is $V * D * 4$ bytes. 
When $M=16, K=256$, $D=128$ as in our experiments, if we use the product quantization with the hidden dimension $h = 8$,  
the codebook size is 131 MB. 
In the WordPiece English token set for BERT, $V\approx 32K$ and  the space for  document-independent 
embeddings  cost about  16.4 MB.
Thus the space cost of the above two parts is insignificant.

The online space cost of token-based document embeddings is $Z * n * (\frac{M\log_2 K} {8} +2)$ bytes. 
Here  each  contextual token embedding of length $D$ is  encoded into a code of length $M$ and 
the space of each code costs $\log_2 K$ bits. For each document,
we also need to store the IDs of its tokens in order to access document-independent embeddings.
We use  2 bytes per token ID in our evaluation because the BERT dictionary based on
WordPiece~\cite{wordpiece} tokenizer has about 32,000 tokens.



In comparison, the
space for  document embeddings in ColBERT  with 2 bytes per number costs   $Z * D * n * 2$ bytes. 
Then the space ratio of ColBERT without CQ and with CQ is about
$\frac{2D \times 8}{M \log_2 K + 2\times 8}$, which is about 14:1 when $D=128$, $M=16$ and $K=256$.
BECR uses 5 layers of the refinement outcome with the BERT encoder for each token 
and stores each layer of the embedding with a 256 bit LSH signature. 
Thus the space cost ratio of  BECR over ColBERT-CQ is approximately 
$\frac{5\times 256}{M \log_2 K + 2\times 8}$, which is about  9:1 when $M=16$ and $K=256$.
We can adjust the parameters of  each of ColBERT, BECR, and ColBERT-CQ for a smaller space with a degraded relevance,
and their space ratio to CQ remains large, which will be  discussed in Section~\ref{sec:eval}.

\label{sec:method} 

\textbf{Time cost for online decompression and composition.}
Let $k$ be the number of documents to re-rank.
The cost of decompression with the short code of a token using the cookbooks is $O(M*h)$ for a product quantizer
and $O(M*D)$ for an additive quantizer. Notice $M*h=D$.
For a one-layer feed-forward network as a composition to recover the final embedding,
the total time cost for decompression and composition is
$O(\text{k}*n*D^2)$ with a product quantizer, and 
$O(\text{k}*n( M*D +D^2))$ with an additive quantizer. 
When using two hidden layers with $D$ dimensions in the first layer output, there is some extra time cost but
the order of time complexity remains unchanged.

\comments{

For two hidden layers, we denote the first layer output dimension as $d_x$,
the number of documents to re-rank as  $k$, and the average number of tokens in a document as $n$. 
The complexity of a product quantizer with two feed-forward layers is $O(\text{k} * n*(h*d_x))$. 
The complexity for an additive quantizer is similar.
In our experiment, $M$ is 16 and $D=128$, $d_x = 128$, and  there is no 
significant difference among these  four architectures. 
}

Noted that because of using  feed-forward layers in final recovery,  our contextual  quantizer
cannot take advantage of an efficiency optimization called asymmetric distance computation in \citet{Jgou2011PQ}.
Since embedding recovery is only applied to top $k$ documents after the first-stage retrieval,
the time efficiency for re-ranking is still reasonable without such an optimization.


\section{Experiments and Evaluation Results}
\label{sec:eval}
\subsection{Settings}

\begin{table}[h]
	\centering
		\resizebox{1.02\columnwidth}{!}{ 
		\begin{tabular}{r r r r r }
			\hline
			\textbf{Dataset} & $\bold{\#}$ \textbf{Query} & $\bold{\#}$ \textbf{Doc} & \textbf{Mean Doc} &  $\bold{\#}$ 
\textbf{Judgments} \\
            & & & \textbf{Length}  & \textbf{ per query} \\
			\hline
			MS MARCO passage Dev & 6980 & 8.8M & 67.5 & 1 \\
			TREC DL 19 passage & 200 & -- & -- & 21 \\
			TREC DL 20 passage & 200 & -- & -- & 18 \\
			\hline
			MS MARCO doc Dev & 5193 & 3.2M & 1460 & 1 \\
			TREC DL 19 doc & 200 & -- & -- & 33 \\
			\hline
		\end{tabular}
	}

	\caption{Dataset statistics. Mean doc length is the average number of WordPiece~\cite{wordpiece} tokens.}

	\label{tab:data}
\end{table}

 
\textbf{Datasets and metrics.} The well-known MS MARCO passage and document ranking datasets are used.
As summarized the in Table~\ref{tab:data}, our evaluation uses
the MS MARCO document and passage collections for document and passage ranking~\cite{Craswell2020OverviewOT, Campos2016MSMARCO}.
The original document and passage ranking tasks  provide 367,013 and 502,940 training queries respectively, 
with about  one judgment label per query. The development query sets are used for relevance evaluation.
The TREC Deep Learning (DL) 2019 and 2020 tracks provide 200 test queries with many judgment labels per query for each task.
\comments{
For document re-ranking, each MS MARCO document is divided into overlapped passage segements of size up to 400 tokens,
and there are  60 overlapping tokens included between two consecutive passage segmentss,  following the ColBERT setup.  
As a result, the number of tokens per document is about 2031 and about 32\% of them  are  contextual overlapping tokens. 
}

Following the official leader-board standard, for the development sets, 
we report mean reciprocal rank (MRR@10, MRR@100) for relevance instead of 
using normalized discounted cumulative gain (NDCG)~\cite{NDCG}  because  such a set has about one judgment label per query, which is too sparse to use NDCG.
For TREC DL test sets which have many judgement lables per query, we report the commonly used NDCG@10 score.
We also measure the dominating space need of the embeddings in bytes and 
re-ranking time latency in milliseconds.
To evaluate latency, we uses an Amazon AWS g4dn instance with Intel Cascade Lake CPUs and  an NVIDIA T4 GPU.  

In all tables below that compare relevance,
we perform paired t-test on 95\% confidence levels. 
In Tables~\ref{tab:overall_psg}, \ref{tab:overall_doc}, and  \ref{tab:storage},
we mark the results with `$^{\dag}$' if the compression method result in statistically significant degradation from 
the ColBERT baseline. In Table~\ref{tab:option}, `$^{\dag}$' is marked for numbers with statistically significant degradation from default setting in the first row.

\textbf{Choices of first-stage retrieval models.}
To retrieve top 1,000 results before re-ranking, we consider the standard fast
BM25 method~\citep{Robertson2009BM25}.
We  have also considered  sparse and dense retrievers that outperform BM25.
We have used uniCOIL~\citep{Lin2021unicoil,2021NAACL-Gao-COIL} as an alternative sparse retriever in Table~\ref{tab:overall_psg}
because it achieves a similar level of relevance as end-to-end ColBERT with a dense retriever, and that of other learned sparse 
representations~\citep{Mallia2021deepimpact, 2021SIGIRFormalSPLADE, Formal2021SPLADEV2}.
ColBERT+uniCOIL has 0.369 MRR while end-to-end
ColBERT  has 0.360 MRR on MSMARCO Dev set.
Moreover, retrieval with a sparse representation such as uniCOIL and BM25
normally uses much less computing resources than a dense retriever.
Relevance numbers reported in some of the previous work on  dense retrieval are derived from the exact search as an upper bound of accuracy. 
When non-exact retrieval techniques such as approximate nearest neighbor  or maximum inner product search are used 
on a more affordable  platform for large datasets, 
there is a  visible loss of relevance~\citep{2021Facebook-DrBoost-Lewis}.
It should be emphasized that the first stage model can be done by either a sparse or a dense retrieval, and this does not affect the applicability of CQ for the second stage
as the focus of this paper.

\textbf{Re-ranking models and quantizers compared.}
We demonstrate the use of CQ for token compression in ColBERT in this paper. 
We compare its relevance with ColBERT, BECR and PreTTR. 
We chose to apply CQ to ColBERT because assuming embeddings are in memory, ColBERT is one of the fastest recent online re-ranking algorithms with strong relevance scores and CQ addresses its embedding storage weakness.
Other re-ranking models compared include:
BERT-base~\citep{Devlin2019BERT}, a cross encoder re-ranker, which takes a query and a document at run time 
and  uses the last layers output from the BERT [CLS] token to generate a ranking score;
TILDEv2~\citep{Zhuang2021TILDEv2}, which expands each document and additively aggregates precomputed neural scores.

We also evaluate the use of unsupervised quantization methods discussed in Section~\ref{sect:background} for ColBERT, 
including two product quantizers (PQ and  OPQ),
and two additive quantizers (RQ and LSQ).


Appendix~\ref{sec:appendix}
 has  additional details  on the retrievers considered, re-ranker implementation, training, and relevance numbers cited. 

\begin{table}[h]
	\centering
		\resizebox{1.02\columnwidth}{!}{
		\begin{tabular}{l | l l l }
			\hline
			Model Specs. & Dev & TREC DL19 & TREC DL20 \\
			& MRR@10 & NDCG@10 & NDCG@10 \\
			\hline
			& \multicolumn{3}{c}{Retrieval choices}\\
			BM25 & 0.172 & 0.425 & 0.453\\
			docT5query & 0.259 & 0.590 & 0.597 \\
			DeepCT$^*$ & 0.243 & 0.572 & -- \\
			TCT-ColBERT(v2) & 0.358 & -- & -- \\
			JPQ$^*$ & 0.341 & 0.677 & -- \\
			DeepImpact  & 0.328 & 0.695 & 0.628 \\
			uniCOIL &  0.347 & 0.703 & 0.675 \\
			\hline
			& \multicolumn{3}{c}{Re-ranking baselines ( +BM25 retrieval)}\\
			BERT-base & 0.349 & 0.682 & 0.655 \\
			BECR & 0.323 & 0.682 & 0.655 \\
			TILDEv2$^*$ & 0.333 & 0.676 & 0.686 \\
			ColBERT  & 0.355 & 0.701 & 0.723 \\
			\hline
			& \multicolumn{3}{c}{Quantization ( +BM25 retrieval)}\\
			ColBERT-PQ & 0.290$^{\dag}$ (-18.3\%) & 0.684 (-2.3\%) & 0.714 (-1.2\%)\\
			ColBERT-OPQ & 0.324$^{\dag}$ (-8.7\%)
			& 0.691 (-1.4\%) & 0.688$^{\dag}$ (-4.8\%) \\
			ColBERT-RQ & -- & 0.675$^{\dag}$ (-3.7\%) & 0.696 (-3.7\%) \\
			ColBERT-LSQ & -- & 0.664$^{\dag}$ (-5.3\%) & 0.656$^{\dag}$ (-9.3\%) \\
			\hline 
			ColBERT-CQ & 0.352 (-0.8\%) & 0.704 (+0.4\%) & 0.716 (-1.0\%) \\
			\hline\hline
			& \multicolumn{3}{c}{ ( +uniCOIL retrieval)}\\
			ColBERT & 0.369 & 0.692 & 0.701 \\
			ColBERT-CQ & 0.360$^{\dag}$ (-2.4\%) & 0.696 (+0.6\%) & 0.720 (+2.7\%) \\
			\hline
		\end{tabular}
		}
	\caption{Relevance scores for MS MARCO passage ranking. 
The \% degradation from ColBERT is listed and  `$^{\dag}$' is marked for statistically significant drop. 
}
	\label{tab:overall_psg}
\end{table}

\begin{table}[h]
	\centering
		\resizebox{0.85\columnwidth}{!}{
		\begin{tabular}{l | l l}
			\hline
			Model Specs. & Dev & TREC DL19\\
			& MRR@100 & NDCG@10 \\
			\hline
			\multicolumn{3}{c}{Retrieval choices}\\
			BM25 & 0.203 & 0.446\\
			docT5query & 0.289 & 0.569 \\
			DeepCT$^*$ & 0.320 & 0.544 \\
			TCT-ColBERT(v2) & 0.351 & -- \\
			JPQ$^*$ & 0.401 & 0.623 \\
			uniCOIL & 0.343 & 0.641 \\
			\hline
			\multicolumn{3}{c}{Re-ranking baselines ( +BM25 retrieval)}\\
			BERT-base$^*$ & 0.393 & 0.670 \\
			ColBERT & 0.410 & 0.714 \\
			\hline
			\multicolumn{3}{c}{Quantization ( +BM25 retrieval)}\\
			ColBERT-PQ & 0.400$^{\dag}$ (-2.4\%) & 0.702 (-1.7\%) \\
			ColBERT-OPQ & 0.404$^{\dag}$ (-1.5\%) & 0.704 (-1.4\%) \\
			ColBERT-RQ & -- & 0.704  (-1.4\%) \\
			ColBERT-LSQ & -- & 0.707 (-1.0\%)  \\
			\hline
			ColBERT-CQ & 0.405$^{\dag}$ (-1.2\%) &  0.712 (-0.3\%) \\
			\hline
		\end{tabular}
		}
	\caption{Relevance scores for MS MARCO document ranking. 
The \% degradation from ColBERT is listed and  `$^{\dag}$' is marked for statistically significant drop.
}
	\label{tab:overall_doc}
\end{table}

\subsection{A Comparison of Relevance}

Table~\ref{tab:overall_psg} and Table~\ref{tab:overall_doc}  show  the ranking relevance in NDCG and MRR  of the 
different
methods and compare against  the use of CQ with ColBERT (marked as ColBERT-CQ).
We either report our experiment results  or cite the relevance numbers from other papers with a $^*$ mark
for such a model. 
For quantization approaches, we adopt M=16, K=256, i.e. compression ratio 14:1 compared to ColBERT.

\comments{
In  Table~\ref{tab:overall_psg}, we choose two retrieval options (BM25 or uniCOIL) to select top 1,000 results for re-ranking. 
BM25 is the standard reference point, which is also the fastest method.
UniCOIL delivers good relevance score compared to other retrievers. As a sparse retriever without the need of GPU,   it is also faster
than dense retrievers in general. 
}

For the passage task, ColBERT outperforms other re-rankers in relevance for  the tested cases.
ColBERT-CQ  after BM25 or uniCOIL retrieval only has a small 
relevance degradation with around 1\% or less, while only requiring 3\% of the storage of ColBERT. 
The relevance of  the ColBERT-CQ+uniCOIL combination is  also competitive to the one
reported in~\citet{Mallia2021deepimpact} for the  ColBERT+DeepImpact combination which has
MRR 0.362 for the Dev query set, NDCG@10 0.722 for TREC DL 2019 and 0.691 for TREC DL 2020.

For the document re-ranking task, Table~\ref{tab:overall_doc}  similarly confirms  the effectiveness 
of ColBERT-CQ. ColBERT-CQ and ColBERT after BM25 retrieval also perform well in general compared to the relevance results of
the  other baselines.
 
\begin{figure}[h!]
    \centering
    \begin{subfigure}{0.23\textwidth}
    \includegraphics[scale=0.14]{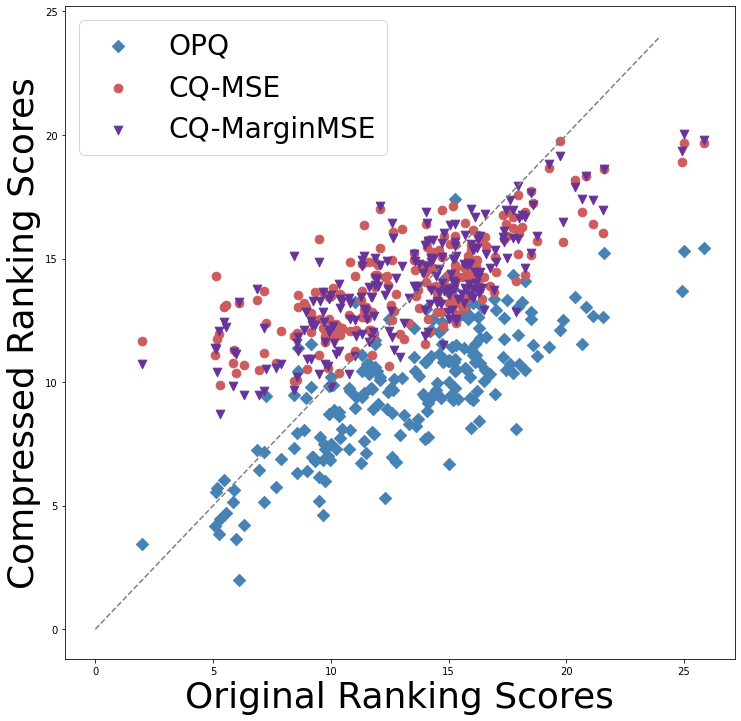}
    \caption{}
    \end{subfigure}%
    \begin{subfigure}{0.2\textwidth}
    \includegraphics[scale=0.14]{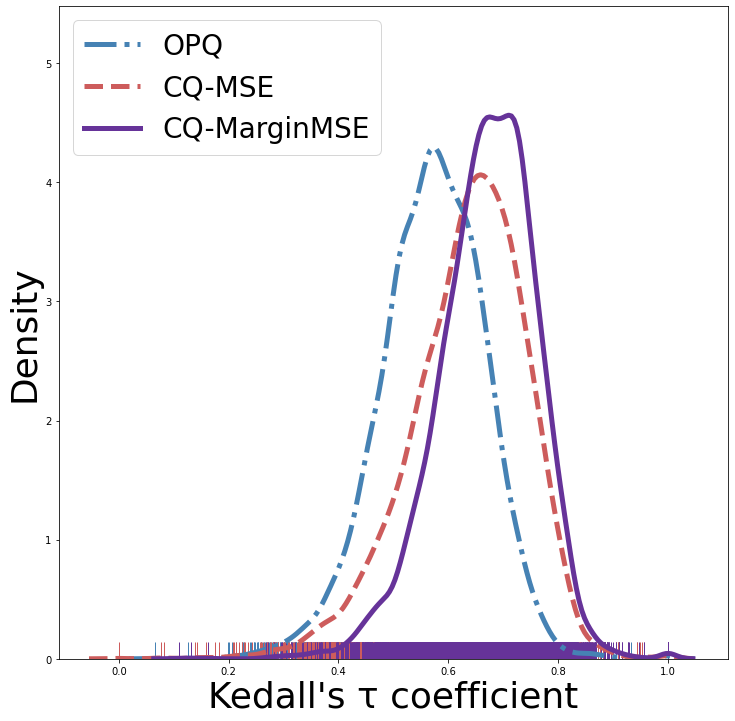}
    \caption{}
    \end{subfigure}
    
    \caption{(a) Ranking score by quantized ColBERT with OPQ and CQ using two loss functions, vs. original ColBERT score. (b) Distribution of Kendall's $\tau$ correlation coefficient between the 1,000 ranked results of quantized and original ColBERT.}
    \label{fig:comp}
\end{figure}

From both Table~\ref{tab:overall_psg} and Table~\ref{tab:overall_doc},  
we observe that in general, CQ significantly outperforms the other quantization approaches (PQ, OPQ, RQ, and LSQ).
As an example, we further explain this by plotting the ranking score of ColBERT with and without 
a quantizer in Figure~\ref{fig:comp}(a).
Compared to OPQ, CQ trained with two  loss functions generates ranking scores 
much closer to the original ColBERT ranking score, and this is also reflected in   Kendall's $\tau$ correlation coefficients  of top 1,000
re-ranked results between a quantized ColBERT  and the original ColBERT (Figure~\ref{fig:comp}(b)). 
There are two reasons that CQ outperforms the other quantizers: 1) The previous quantizers
do not perform contextual decomposition to isolate intrinsic context-independent information in embeddings, and thus their approximation
yields more relevance loss; 
2) Their training loss function is not tailored to the re-ranking task.

\subsection{Effectiveness on Space Reduction}

\begin{table}[h]
	\centering
		\resizebox{1.02 \columnwidth}{!}{
		\begin{tabular}{r| r | r r r r}
			\hline
			 & \textbf{Doc task} & \multicolumn{4}{|c}{\textbf{Passage task}} \\
			\textbf{Model} & \textbf{Space} & \textbf{Space} & \textbf{Disk I/O} &  \textbf{Latency } &  \textbf{MRR@10} \\
			\hline
			BECR &  791G& 89.9G &  -- &  8ms & 0.323 \\
			PreTTR$^*$ & -- & 2.6T & >182ms & >1000ms & 0.358\\
			TILDEv2$^*$ & -- & 5.2G & -- & -- & 0.326 \\
			ColBERT & 1.6T & 143G &>182ms  & 16ms & 0.355 \\
			ColBERT-small$^*$ & 300G & 26.8G  & -- & -- & 0.339\\
			ColBERT-OPQ & 112G & 10.2G & -- &  56ms & 0.324$^{\dag}$ \\
			\hline
			ColBERT-CQ & & & \\
			undecomposed & 112G & 10.2G & -- & 17ms & 0.339$^{\dag}$ \\
			K=256 &112G & 10.2G & --& 17ms & 0.352\\
			K=16 & 62G & 5.6G & --&  17ms & 0.339$^{\dag}$\\
			K=4 & 37G & 3.4G & --& 17ms & 0.326$^{\dag}$ \\
			\hline
		\end{tabular}
		}
	\caption{Embedding space size in bytes for the document ranking task and  for
 the passage ranking task.   Re-ranking time per query and relevance for top 1,000 passages
in milliseconds on a GPU using the Dev query set. M=16. 
For ColBERT-OPQ and ColBERT-CQ-undecomposed, K=256.
Assume passage embeddings in PreTTR and ColBERT do not fit in memory.
`$^{\dag}$' is marked for MRR numbers  with statistically significant degradation from the ColBERT baseline. 
}
	\label{tab:storage}
\end{table}

Table~\ref{tab:storage} shows the estimated space size in bytes for embeddings in the MS MARCO document and passage corpora, and compares CQ with other approaches.  
Each MS MARCO document is divided into overlapped passage segments of size up to about 400 tokens,
and there are  60 tokens overlapped between two consecutive passage segments,  following the ColBERT setup.
As a result, the average number of WordPiece tokens per document changes from 1460 to  about 2031 due to  the addition of  overlapping contextual tokens.

\comments{
Document storage is high because we split each document into passages of size 400 tokens with 
60 overlapping tokens, following the ColBERT setup. 
}
To demonstrate the tradeoff, 
we also list their  estimated time latency and relevance  in passage re-ranking as a reference
and notice that more relevance comparison  results are in Tables~\ref{tab:overall_psg} and ~\ref{tab:overall_doc}.
The latency is the total time for embedding decompression/recovery and re-ranking. 

For PreTTR and ColBERT, we assume that their passage embedding data cannot fit in memory given their large data sizes.
The disk I/O latency number is based on their passage embedding size and our  test on
a Samsung 870 QVO solid-state disk drive to fetch 1,000 passage embeddings randomly. 
Their I/O latency takes about 
182ms with single-thread I/O and with no I/O contention, 
and their disk access can incur  much  more time when multiple queries are processed in parallel in a server 
dealing with many clients.  For example, fetching 1,000 passage embeddings for each of ColBERT and PreTTR
takes about 1,001ms and 3,870ms respectively when the server is handling 16 and 64 queries simultaneously with multiple threads.
%

For other methods, their passage embedding data is relatively small and we assume that it can be preloaded in memory.
The query latency reported in the 4-th column of 
Table~\ref{tab:storage} 
excludes the 
first-stage retrieval time.  The default ColBERT uses embedding dimension 128 and 2 byte floating numbers. 
ColBERT-small denotes an optional configuration  suggested from the ColBERT paper using 24 embedding dimensions and 2-byte floating numbers
with a degraded relevance performance. 

As shown in Table~\ref{tab:storage}, the embedding footprint of ColBERT CQ uses 
about  112GB and 10.2GB, 
respectively for document and passage re-ranking tasks. 
By looking at the latency difference of ColBERT with and without CQ, the time overhead of CQ
for decompression and embedding recovery takes 1ms per query, which is insignificant.

Compared with another quantizer ColBERT-OPQ, 
ColBERT-CQ can achieve the same level of space saving with $K=256$ while having a substantial relevance improvement.
ColBERT-CQ with $K=4$ achieves the same level of relevance as ColBERT-OPQ while  yielding a storage reduction of 67\% and  a latency 
reduction of about 70\%. 
Comparing ColBERT-CQ with no contextual decomposition, under the same space cost, ColBERT-CQ's relevance is 4\% higher.
CQ with $K=16$ achieves the same level relevance as ColBERT-CQ-undecomposed with $K=256$, while the storage of CQ reduces by 44\%.
Comparing with ColBERT-small which adopts more aggressive space reduction, 
ColBERT-CQ with $K=16$ would be competitive in relevance while its space is about 4x smaller.

Comparing with other non-ColBERT baselines (BECR, PreTTR, and TILDEv2), ColBERT-CQ strikes a good balance across relevance, space and latency. 
For the fast CPU based model (BECR, TILDEv2), our model achieves better relevance with either lower or comparable space usage. 
For BECR,  its embedding footprint  with 89.9GB  may fit in memory for MS MARCO  passages, 
it becomes very  expensive to configure  a machine with  much more memory  for  BECR's MS MARCO document embeddings with about 791GB.

\subsection{Design Options for CQ}
\begin{table}[h]
	\centering
		\resizebox{1.0\columnwidth}{!}{
		\begin{tabular}{l r r}
			\hline\hline
			 & \textbf{TREC19} &  \textbf{TREC20} \\
			\hline
			CQ, Product, 1 layer, MarginMSE & 0.687& 0.713\\
			\hline \hline
			\textbf{Different model configurations} & & \\
			No decomposition. Product  & 0.663$^{\dag}$	& 0.686 \\
			No decomposition. Additive  & 0.656$^{\dag}$ &	0.693 \\
			CQ, Product, 1 layer, & & \\
			\hspace{1cm} raw static embedding  & 0.655$^{\dag}$ & 0.683$^{\dag}$\\
			CQ, Additive, 1 layer & 0.693 & 0.703\\
			CQ, Product, 2 layers & 0.683 & 0.707\\
			CQ, Additive, 2 layers & 0.688 & 0.703\\
			\hline \hline
			\textbf{Different training loss functions} &  &  \\
			CQ, Product, 1 layer, MSE & 0.679 &	0.704\\
			CQ, Product, 1 layer, PairwiseCE & 0.683 & 0.705\\
			\hline
		\end{tabular}
		}
	\caption{NDCG@10 of different design options for CQ  
in TREC DL passage ranking. If the compression method result in statistically significant degradation from the default setting, `$^{\dag}$' is marked.
} 

	\label{tab:option}
\end{table}

Table~\ref{tab:option} 
shows the relevance scores for the TREC deep learning passage ranking task with different design options for CQ. 
As an alternative setting,  the codebooks in this table use M=16 and  K=32 with  compression ratio 21:1  compared to ColBERT. 
Row 1 is the default design configuration for CQ with product operators and 1 composition layer, and the MarginMSE loss function. 

{\bf Different architecture or quantization options.}
Rows 2 and 3 of Table~\ref{tab:option} 
denote CQ using product or additive operators without decomposing each embedding into two components,
and there is about 4\% degradation without such decomposition.

Row 4 changes CQ using the raw static embeddings of tokens from BERT instead of the upper layer outcome of BERT encoder
and there is an up to 4.7\% degradation. Notice such a strategy is used in SDR. 
From Row 5 to Row 7, we change CQ to use additive operators or use a two-layer composition. 
The performance of  product or additive operators is in a similar level while the benefit of using two layers is relatively small.

{\bf Different training loss functions for CQ.}
Last two rows  of Table~\ref{tab:option} 
use the MSE and PairwiseCE loss functions, respectively.
There is an about 1.2\% improvement using MarginMSE.
Figure~\ref{fig:comp} gives an explanation why MarginMSE is more effective. 
While CQ trained with MSE and MarginMSE generates ranking scores 
close to the original ranking scores in Figure~\ref{fig:comp}(a),
the distribution of  Kendall's $\tau$ correlation coefficients of 1,000 passages
in Figure~\ref{fig:comp}(b) shows that the passage rank order derived by CQ with the MarginMSE loss has a better correlation with 
that by ColBERT.

\section{Concluding Remarks}

Our evaluation shows the effectiveness of CQ used for ColBERT in 
compressing the space of token  embeddings with about 14:1 ratio  
while incurring a small relevance degradation
in MS MARCO passage and document re-ranking tasks. The quantized token-based document  embeddings for the tested cases 
can be hosted in memory for fast and high-throughput access. 
This is accomplished by a neural network that decomposes ranking contributions of contextual embeddings, and  
jointly trains context-aware  decomposition and quantization with a loss 
function preserving ranking accuracy.
The online time cost to decompress and recover embeddings is insignificant with 1ms for the tested cases. 
The CQ implementation is available at https://github.com/yingrui-yang/ContextualQuantizer.

\begin{table}[h]
        \centering
                \begin{small}
                \begin{tabular}{r | r| r }
                        \hline
                        Model & ColBERTv2 retri.&  uniCOIL-ColBERTv2-CQ  \\
                        \hline
                        Space & 29GB (2.8x)  &  10.2GB (1x)\\
                        \hline
                       Dev & 0.397 MRR@10 &  0.387 (-2.5\%) \\
		       	DL 19  & -- & 0.746  NDCG@10 \\
			DL 20  & -- &  0.726  NDCG@10 \\
                        \hline
                \end{tabular}
                \end{small}
                
        \caption{ ColBERTv2/CQ for MS MARCO passages }
        \label{tab:colBERTv2}
\end{table}
CQ is also applicable to re-ranking with the
contemporaneous work ColBERTv2~\cite{Santhanam2021ColBERTv2}. 
While ColBERTv2 is proposed for dense retrieval, 
we use its uncompressed model for ColBERT re-ranking and compress its embeddings with CQ (M=16, K=256)
after getting top 1,000 with uniCOIL sparse retrieval.  Table~\ref{tab:colBERTv2} lists the results for MS MARCO passages.
The above setting has  a -2.5\% MRR@10 loss for the Dev set
compared to the number reported in ColBERTv2 as a standalone retriever. 
Such dense retrieval with multi-vector representation is likely to be much more expensive than sparse retrieval on a large dataset.
%
Thus the relevance of ColBERTv2+CQ for re-ranking with uniCOIL sparse retrieval is fairly strong, achievable  with a reasonable latency and limited
computing resource while its  embedding footprint is 2.8x smaller than that  of  standalone ColBERTv2.
Our future work is to investigate the above issue further and study the use of CQ in other re-ranking methods.

{\bf Acknowledgments}. We thank Cindy Zhao, Jiahua Wang,  and anonymous referees
for their valuable comments and/or help. 
This work is supported in part by NSF IIS-2040146  and by a Google faculty research award.
It has used the Extreme Science and Engineering Discovery Environment
supported by NSF ACI-1548562.
Any opinions, findings, conclusions or recommendations expressed in this material
are those of the authors and do not necessarily reflect the views of the NSF.




\bibliography{compression,comprank,ranking,oldranking,custom,mise}

\begin{thebibliography}{58}
\expandafter\ifx\csname natexlab\endcsname\relax\def\natexlab#1{#1}\fi

\bibitem[{Ai et~al.(2015)Ai, Yu, Wu, He, and Guan}]{Ai2015RQ}
Liefu Ai, Junqing Yu, Zenbin Wu, Yunfeng He, and Tao Guan. 2015.
\newblock Optimized residual vector quantization for efficient approximate
  nearest neighbor search.
\newblock \emph{Multimedia Systems}, 23:169--181.

\bibitem[{Campos et~al.(2016)Campos, Nguyen, Rosenberg, Song, Gao, Tiwary,
  Majumder, Deng, and Mitra}]{Campos2016MSMARCO}
Daniel~Fernando Campos, Tri Nguyen, Mir Rosenberg, Xia Song, Jianfeng Gao,
  Saurabh Tiwary, Rangan Majumder, Li~Deng, and Bhaskar Mitra. 2016.
\newblock Ms marco: A human generated machine reading comprehension dataset.
\newblock \emph{ArXiv}, abs/1611.09268.

\bibitem[{Chen et~al.(2020{\natexlab{a}})Chen, Yang, Raman, Bendersky, Yeh,
  Zhou, Najork, Cai, and Emadzadeh}]{Chen2020DiPairFA}
Jiecao Chen, Liu Yang, Karthik Raman, Michael Bendersky, Jung-Jung Yeh, Yun
  Zhou, Marc Najork, D.~Cai, and Ehsan Emadzadeh. 2020{\natexlab{a}}.
\newblock Dipair: Fast and accurate distillation for trillion-scale text
  matching and pair modeling.
\newblock In \emph{EMNLP}.

\bibitem[{Chen et~al.(2020{\natexlab{b}})Chen, He, Hui, Sun, and
  Sun}]{Chen2020-SimplifiedTK}
Xuanang Chen, B.~He, Kai Hui, L.~Sun, and Yingfei Sun. 2020{\natexlab{b}}.
\newblock Simplified tinybert: Knowledge distillation for document retrieval.
\newblock \emph{ArXiv}, abs/2009.07531.

\bibitem[{Cohen et~al.(2021)Cohen, Portnoy, Fetahu, and
  Ingber}]{2021-SDR-Cohen}
Nachshon Cohen, Amit Portnoy, Besnik Fetahu, and Amir Ingber. 2021.
\newblock Sdr: Efficient neural re-ranking using succinct document
  representation.
\newblock \emph{ArXiv}, 2110.02065.

\bibitem[{Craswell et~al.(2020)Craswell, Mitra, Yilmaz, Campos, and
  Voorhees}]{Craswell2020OverviewOT}
Nick Craswell, Bhaskar Mitra, Emine Yilmaz, Daniel~Fernando Campos, and
  Ellen~M. Voorhees. 2020.
\newblock Overview of the trec 2020 deep learning track.
\newblock \emph{ArXiv}, abs/2102.07662.

\bibitem[{Dai and Callan(2019)}]{bertmaxp}
Zhuyun Dai and J.~Callan. 2019.
\newblock Deeper text understanding for ir with contextual neural language
  modeling.
\newblock \emph{Proceedings of the 42nd International ACM SIGIR Conference on
  Research and Development in Information Retrieval}.

\bibitem[{Dai and Callan(2020)}]{Dai2020deepct}
Zhuyun Dai and Jamie Callan. 2020.
\newblock Context-aware term weighting for first stage passage retrieval.
\newblock \emph{SIGIR}.

\bibitem[{Dai et~al.(2018)Dai, Xiong, Callan, and Liu}]{convknrm}
Zhuyun Dai, Chenyan Xiong, Jamie Callan, and Zhiyuan Liu. 2018.
\newblock Convolutional neural networks for soft-matching n-grams in ad-hoc
  search.
\newblock In \emph{WSDM}, pages 126--134.

\bibitem[{Devlin et~al.(2019)Devlin, Chang, Lee, and
  Toutanova}]{Devlin2019BERT}
Jacob Devlin, Ming-Wei Chang, Kenton Lee, and Kristina Toutanova. 2019.
\newblock Bert: Pre-training of deep bidirectional transformers for language
  understanding.
\newblock In \emph{NAACL}.

\bibitem[{Formal et~al.(2021{\natexlab{a}})Formal, Lassance, Piwowarski, and
  Clinchant}]{Formal2021SPLADEV2}
Thibault Formal, C.~Lassance, Benjamin Piwowarski, and St{\'e}phane Clinchant.
  2021{\natexlab{a}}.
\newblock Splade v2: Sparse lexical and expansion model for information
  retrieval.
\newblock \emph{ArXiv}, abs/2109.10086.

\bibitem[{Formal et~al.(2021{\natexlab{b}})Formal, Piwowarski, and
  Clinchant}]{2021SIGIRFormalSPLADE}
Thibault Formal, Benjamin Piwowarski, and St\'{e}phane Clinchant.
  2021{\natexlab{b}}.
\newblock \emph{SPLADE: Sparse Lexical and Expansion Model for First Stage
  Ranking}, pages 2288--2292. ACM.

\bibitem[{Gao and Callan(2021)}]{gao-2021-condenser}
Luyu Gao and Jamie Callan. 2021.
\newblock \href {https://doi.org/10.18653/v1/2021.emnlp-main.75} {Condenser: a
  pre-training architecture for dense retrieval}.
\newblock In \emph{Proceedings of the 2021 Conference on Empirical Methods in
  Natural Language Processing}, pages 981--993, Online and Punta Cana,
  Dominican Republic. Association for Computational Linguistics.

\bibitem[{Gao et~al.(2020)Gao, Dai, and Callan}]{Gao2020distill}
Luyu Gao, Zhuyun Dai, and J.~Callan. 2020.
\newblock Understanding bert rankers under distillation.
\newblock \emph{Proceedings of SIGIR}.

\bibitem[{Gao et~al.(2021)Gao, Dai, and Callan}]{2021NAACL-Gao-COIL}
Luyu Gao, Zhuyun Dai, and Jamie Callan. 2021.
\newblock {COIL:} revisit exact lexical match in information retrieval with
  contextualized inverted list.
\newblock \emph{NAACL}.

\bibitem[{Ge et~al.(2013)Ge, He, Ke, and Sun}]{Ge2013OPQ}
Tiezheng Ge, Kaiming He, Qifa Ke, and Jian Sun. 2013.
\newblock Optimized product quantization for approximate nearest neighbor
  search.
\newblock \emph{CVPR}, pages 2946--2953.

\bibitem[{Guo et~al.(2016)Guo, Fan, Ai, and Croft}]{drmm}
J.~Guo, Y.~Fan, Qingyao Ai, and W.~Croft. 2016.
\newblock A deep relevance matching model for ad-hoc retrieval.
\newblock \emph{CIKM}.

\bibitem[{Hofst{\"a}tter et~al.(2020{\natexlab{a}})Hofst{\"a}tter, Althammer,
  Schr{\"o}der, Sertkan, and Hanbury}]{Hofsttter2020marginMSE}
Sebastian Hofst{\"a}tter, Sophia Althammer, Michael Schr{\"o}der, Mete Sertkan,
  and Allan Hanbury. 2020{\natexlab{a}}.
\newblock Improving efficient neural ranking models with cross-architecture
  knowledge distillation.
\newblock \emph{ArXiv}, abs/2010.02666.

\bibitem[{Hofst{\"a}tter et~al.(2020{\natexlab{b}})Hofst{\"a}tter, Zamani,
  Mitra, Craswell, and Hanbury}]{tkl}
Sebastian Hofst{\"a}tter, Hamed Zamani, Bhaskar Mitra, Nick Craswell, and
  A.~Hanbury. 2020{\natexlab{b}}.
\newblock Local self-attention over long text for efficient document retrieval.
\newblock \emph{SIGIR}.

\bibitem[{Hofst{\"a}tter et~al.(2020{\natexlab{c}})Hofst{\"a}tter, Zlabinger,
  and Hanbury}]{2020ECAI-tk}
Sebastian Hofst{\"a}tter, Markus Zlabinger, and A.~Hanbury. 2020{\natexlab{c}}.
\newblock Interpretable \& time-budget-constrained contextualization for
  re-ranking.
\newblock In \emph{ECAI}.

\bibitem[{Jang et~al.(2017)Jang, Gu, and Poole}]{Jang2017Gumbel}
Eric Jang, Shixiang~Shane Gu, and Ben Poole. 2017.
\newblock Categorical reparameterization with gumbel-softmax.
\newblock \emph{ICLR}.

\bibitem[{J{\"a}rvelin and Kek{\"a}l{\"a}inen(2002)}]{NDCG}
Kalervo J{\"a}rvelin and Jaana Kek{\"a}l{\"a}inen. 2002.
\newblock Cumulated gain-based evaluation of ir techniques.
\newblock \emph{ACM Transactions on Information Systems (TOIS)},
  20(4):422--446.

\bibitem[{J{\'e}gou et~al.(2011)J{\'e}gou, Douze, and Schmid}]{Jgou2011PQ}
Herv{\'e} J{\'e}gou, Matthijs Douze, and Cordelia Schmid. 2011.
\newblock Product quantization for nearest neighbor search.
\newblock \emph{IEEE Transactions on Pattern Analysis and Machine
  Intelligence}, 33:117--128.

\bibitem[{Ji et~al.(2019)Ji, Shao, and Yang}]{JiWWW2019}
Shiyu Ji, Jinjin Shao, and Tao Yang. 2019.
\newblock Efficient interaction-based neural ranking with locality sensitive
  hashing.
\newblock In \emph{WWW}.

\bibitem[{Johnson et~al.(2017)Johnson, Douze, and J{\'e}gou}]{JDH17faiss}
Jeff Johnson, Matthijs Douze, and Herv{\'e} J{\'e}gou. 2017.
\newblock Billion-scale similarity search with gpus.
\newblock \emph{IEEE Transactions on Big Data}.

\bibitem[{Karpukhin et~al.(2020)Karpukhin, Oğuz, Min, Lewis, Wu, Edunov, Chen,
  and tau Yih}]{Karpukhin2020DensePR}
V.~Karpukhin, Barlas Oğuz, Sewon Min, Patrick Lewis, Ledell~Yu Wu, Sergey
  Edunov, Danqi Chen, and Wen tau Yih. 2020.
\newblock Dense passage retrieval for open-domain question answering.
\newblock \emph{ArXiv}, abs/2010.08191.

\bibitem[{Khattab and Zaharia(2020)}]{colbert}
O.~Khattab and M.~Zaharia. 2020.
\newblock Colbert: Efficient and effective passage search via contextualized
  late interaction over bert.
\newblock \emph{SIGIR}.

\bibitem[{Kingma and Ba(2015)}]{Kingma2015Adam}
Diederik~P. Kingma and Jimmy Ba. 2015.
\newblock Adam: A method for stochastic optimization.
\newblock \emph{CoRR}, abs/1412.6980.

\bibitem[{Lewis et~al.(2021)Lewis, Oğuz, Xiong, Petroni, tau Yih, and
  Riedel}]{2021Facebook-DrBoost-Lewis}
Patrick Lewis, Barlas Oğuz, Wenhan Xiong, Fabio Petroni, Wen tau Yih, and
  Sebastian Riedel. 2021.
\newblock \href {http://arxiv.org/abs/2112.07771} {Boosted dense retriever}.

\bibitem[{Li et~al.(2020)Li, Yates, MacAvaney, He, and Sun}]{parade}
Canjia Li, A.~Yates, Sean MacAvaney, B.~He, and Yingfei Sun. 2020.
\newblock Parade: Passage representation aggregation for document reranking.
\newblock \emph{ArXiv}, abs/2008.09093.

\bibitem[{Lin et~al.(2020)Lin, Nogueira, and Yates}]{Lin2020PretrainedTF}
Jimmy Lin, Rodrigo Nogueira, and A.~Yates. 2020.
\newblock Pretrained transformers for text ranking: Bert and beyond.
\newblock \emph{ArXiv}, abs/2010.06467.

\bibitem[{Lin and Ma(2021)}]{Lin2021unicoil}
Jimmy~J. Lin and Xueguang Ma. 2021.
\newblock A few brief notes on deepimpact, coil, and a conceptual framework for
  information retrieval techniques.
\newblock \emph{ArXiv}, abs/2106.14807.

\bibitem[{Lin et~al.(2021)Lin, Yang, and Lin}]{Lin2021tctcolbert}
Sheng-Chieh Lin, Jheng-Hong Yang, and Jimmy~J. Lin. 2021.
\newblock In-batch negatives for knowledge distillation with tightly-coupled
  teachers for dense retrieval.
\newblock In \emph{REPL4NLP}.

\bibitem[{MacAvaney et~al.(2020)MacAvaney, Nardini, Perego, Tonellotto,
  Goharian, and Frieder}]{MacAvaney2020SIGIR-prettr}
Sean MacAvaney, F.~Nardini, R.~Perego, N.~Tonellotto, Nazli Goharian, and
  O.~Frieder. 2020.
\newblock Efficient document re-ranking for transformers by precomputing term
  representations.
\newblock \emph{SIGIR}.

\bibitem[{MacAvaney et~al.(2019)MacAvaney, Yates, Cohan, and
  Goharian}]{MacAvaney2019SIGIR-cedr}
Sean MacAvaney, Andrew Yates, Arman Cohan, and Nazli Goharian. 2019.
\newblock Cedr: Contextualized embeddings for document ranking.
\newblock \emph{SIGIR}.

\bibitem[{Maddison et~al.(2017)Maddison, Mnih, and Teh}]{Maddison2017Gumbel}
Chris~J. Maddison, Andriy Mnih, and Yee~Whye Teh. 2017.
\newblock The concrete distribution: A continuous relaxation of discrete random
  variables.
\newblock \emph{ICLR}.

\bibitem[{Mallia et~al.(2021)Mallia, Khattab, Tonellotto, and
  Suel}]{Mallia2021deepimpact}
Antonio Mallia, O.~Khattab, Nicola Tonellotto, and Torsten Suel. 2021.
\newblock Learning passage impacts for inverted indexes.
\newblock \emph{SIGIR}.

\bibitem[{Martinez et~al.(2018)Martinez, Zakhmi, Hoos, and
  Little}]{Martinez2018ECCV-LSQ}
Julieta Martinez, Shobhit Zakhmi, Holger~H. Hoos, and J.~Little. 2018.
\newblock Lsq++: Lower running time and higher recall in multi-codebook
  quantization.
\newblock In \emph{ECCV}.

\bibitem[{Mitra et~al.(2021)Mitra, Hofst{\"a}tter, Zamani, and
  Craswell}]{2021SIGIR-ck-Mitra}
Bhaskar Mitra, Sebastian Hofst{\"a}tter, Hamed Zamani, and Nick Craswell. 2021.
\newblock Conformer-kernel with query term independence for document retrieval.
\newblock \emph{SIGIR}.

\bibitem[{Nogueira and Cho(2019)}]{Nogueira2019PassageRW}
Rodrigo Nogueira and Kyunghyun Cho. 2019.
\newblock Passage re-ranking with bert.
\newblock \emph{ArXiv}, abs/1901.04085.

\bibitem[{Nogueira et~al.(2019{\natexlab{a}})Nogueira, Yang, Cho, and
  Lin}]{monobert}
Rodrigo Nogueira, W.~Yang, Kyunghyun Cho, and Jimmy Lin. 2019{\natexlab{a}}.
\newblock Multi-stage document ranking with bert.
\newblock \emph{ArXiv}, abs/1910.14424.

\bibitem[{Nogueira et~al.(2019{\natexlab{b}})Nogueira, Yang, Lin, and
  Cho}]{Nogueira2019d2q}
Rodrigo Nogueira, Wei Yang, Jimmy~J. Lin, and Kyunghyun Cho.
  2019{\natexlab{b}}.
\newblock Document expansion by query prediction.
\newblock \emph{ArXiv}, abs/1904.08375.

\bibitem[{Reimers and Gurevych(2019)}]{Reimers2019SentenceBERTSE}
Nils Reimers and Iryna Gurevych. 2019.
\newblock Sentence-bert: Sentence embeddings using siamese bert-networks.
\newblock In \emph{EMNLP/IJCNLP}.

\bibitem[{Ren et~al.(2021)Ren, Qu, Liu, Zhao, She, Wu, Wang, and
  Wen}]{2021EMNLP-Ren-RocketQAv2}
Ruiyang Ren, Yingqi Qu, Jing Liu, Wayne~Xin Zhao, QiaoQiao She, Hua Wu, Haifeng
  Wang, and Ji-Rong Wen. 2021.
\newblock {R}ocket{QA}v2: A joint training method for dense passage retrieval
  and passage re-ranking.
\newblock In \emph{Proceedings of the 2021 Conference on Empirical Methods in
  Natural Language Processing}, pages 2825--2835, Online and Punta Cana,
  Dominican Republic. Association for Computational Linguistics.

\bibitem[{Robertson and Zaragoza(2009)}]{Robertson2009BM25}
Stephen~E. Robertson and Hugo Zaragoza. 2009.
\newblock The probabilistic relevance framework: Bm25 and beyond.
\newblock \emph{Found. Trends Inf. Retr.}, 3:333--389.

\bibitem[{Santhanam et~al.(2021)Santhanam, Khattab, Saad-Falcon, Potts, and
  Zaharia}]{Santhanam2021ColBERTv2}
Keshav Santhanam, O.~Khattab, Jon Saad-Falcon, Christopher Potts, and Matei~A.
  Zaharia. 2021.
\newblock Colbertv2: Effective and efficient retrieval via lightweight late
  interaction.
\newblock \emph{ArXiv}, abs/2112.01488.

\bibitem[{Shu and Nakayama(2018)}]{Shu2018ICLR}
Raphael Shu and Hideki Nakayama. 2018.
\newblock Compressing word embeddings via deep compositional code learning.
\newblock \emph{ICLR}.

\bibitem[{Wu et~al.(2016)Wu, Schuster, Chen, Le, Norouzi, Macherey, Krikun,
  Cao, Gao, Macherey, Klingner, Shah, Johnson, Liu, Kaiser, Gouws, Kato, Kudo,
  Kazawa, Stevens, Kurian, Patil, Wang, Young, Smith, Riesa, Rudnick, Vinyals,
  Corrado, Hughes, and Dean}]{wordpiece}
Y.~Wu, M.~Schuster, Z.~Chen, Q.~Le, M.~Norouzi, W.~Macherey, M.~Krikun, Y.~Cao,
  Q.~Gao, K.~Macherey, J.~Klingner, A.~Shah, M.~Johnson, X.~Liu, L.~Kaiser,
  S.~Gouws, Y.~Kato, T.~Kudo, H.~Kazawa, K.~Stevens, G.~Kurian, N.~Patil,
  W.~Wang, C.~Young, J.~Smith, J.~Riesa, A.~Rudnick, O.~Vinyals, G.~Corrado,
  M.~Hughes, and J.~Dean. 2016.
\newblock Google's neural machine translation system: Bridging the gap between
  human and machine translation.
\newblock \emph{ArXiv}, abs/1609.08144.

\bibitem[{Xin et~al.(2020)Xin, Nogueira, Yu, and Lin}]{Xin2020SUSTAINLP}
J.~Xin, Rodrigo Nogueira, Y.~Yu, and Jimmy Lin. 2020.
\newblock Early exiting bert for efficient document ranking.
\newblock In \emph{SUSTAINLP}.

\bibitem[{Xiong et~al.(2017)Xiong, Dai, Callan, Liu, and
  Power}]{Xiong2017SIGIR-knrm}
Chenyan Xiong, Zhuyun Dai, J.~Callan, Zhiyuan Liu, and R.~Power. 2017.
\newblock End-to-end neural ad-hoc ranking with kernel pooling.
\newblock \emph{SIGIR}.

\bibitem[{Xiong et~al.(2021)Xiong, Xiong, Li, Tang, Liu, Bennett, Ahmed, and
  Overwijk}]{xiong2021-ANCE}
Lee Xiong, Chenyan Xiong, Ye~Li, Kwok-Fung Tang, Jialin Liu, Paul~N. Bennett,
  Junaid Ahmed, and Arnold Overwijk. 2021.
\newblock \href {https://openreview.net/forum?id=zeFrfgyZln} {Approximate
  nearest neighbor negative contrastive learning for dense text retrieval}.
\newblock In \emph{International Conference on Learning Representations}.

\bibitem[{Yang et~al.(2019)Yang, Zhang, and Lin}]{birch}
Wei Yang, Haotian Zhang, and Jimmy Lin. 2019.
\newblock Simple applications of bert for ad hoc document retrieval.
\newblock \emph{ArXiv}, abs/1903.10972.

\bibitem[{Yang et~al.(2022)Yang, Qiao, Shao, Yan, and Yang}]{Yang2021WSDM-BECR}
Yingrui Yang, Yifan Qiao, Jinjin Shao, Xifeng Yan, and Tao Yang. 2022.
\newblock Lightweight composite re-ranking for efficient keyword search with
  {BERT}.
\newblock \emph{WSDM}.

\bibitem[{Zhan et~al.(2020)Zhan, Mao, Liu, Zhang, and Ma}]{Zhan2020LearningTR}
Jingtao Zhan, J.~Mao, Yiqun Liu, Min Zhang, and Shaoping Ma. 2020.
\newblock Learning to retrieve: How to train a dense retrieval model
  effectively and efficiently.
\newblock \emph{ArXiv}, abs/2010.10469.

\bibitem[{Zhan et~al.(2021{\natexlab{a}})Zhan, Mao, Liu, Guo, Zhang, and
  Ma}]{2021CIKM-JPQ-Zhan}
Jingtao Zhan, Jiaxin Mao, Yiqun Liu, Jiafeng Guo, Min Zhang, and Shaoping Ma.
  2021{\natexlab{a}}.
\newblock Jointly optimizing query encoder and product quantization to improve
  retrieval performance.
\newblock \emph{CIKM}.

\bibitem[{Zhan et~al.(2021{\natexlab{b}})Zhan, Mao, Liu, Guo, Zhang, and
  Ma}]{2021SIGIR-Zhan-ADORE-dense}
Jingtao Zhan, Jiaxin Mao, Yiqun Liu, Jiafeng Guo, Min Zhang, and Shaoping Ma.
  2021{\natexlab{b}}.
\newblock \href {https://arxiv.org/abs/2104.08051} {Optimizing dense retrieval
  model training with hard negatives}.
\newblock \emph{CoRR}, abs/2104.08051.

\bibitem[{Zhang et~al.(2020)Zhang, Nie, Geng, Ramamurthy, Song, and
  Jiang}]{dcbert}
Y.~Zhang, Ping Nie, Xiubo Geng, A.~Ramamurthy, L.~Song, and Daxin Jiang. 2020.
\newblock Dc-bert: Decoupling question and document for efficient contextual
  encoding.
\newblock In \emph{SIGIR}.

\bibitem[{Zhuang and Zuccon(2021)}]{Zhuang2021TILDEv2}
Shengyao Zhuang and G.~Zuccon. 2021.
\newblock Fast passage re-ranking with contextualized exact term matching and
  efficient passage expansion.
\newblock \emph{ArXiv}, abs/2108.08513.

\end{thebibliography}


\bibliographystyle{acl_natbib}

\clearpage
\appendix

\section{Details on Retrieval Choices, Numbers Cited, and Model Implementations}
\label{sec:appendix}

\textbf{First-stage retrieval models considered.}
To retrieve top results before re-ranking,
we  have considered  the recent work in sparse and dense retrieval that outperforms BM25.
For sparse retrieval with inverted indices,
DeepCT~\citep{Dai2020deepct} uses deep learning to assign more sophisticated term
weights for soft matching. The docT5query work~\citep{Nogueira2019d2q}
uses a  neural  model to pre-process and expand documents.
The recent work on sparse representations includes DeepImpact~\citep{Mallia2021deepimpact},  uniCOIL~\citep{Lin2021unicoil,2021NAACL-Gao-COIL},
and SPLADE~\cite{2021SIGIRFormalSPLADE, Formal2021SPLADEV2}, 
for learning neural  contextualized term weights with document expansion.
\comments{
There are a number of schemes proposed for dense retrieval.
For example, TCT-ColBERT(v2)~\citep{Lin2021tctcolbert} is a recent scheme
that produces a dense document representation  with knowledge distillation,
and
JPQ~\citep{2021CIKM-JPQ-Zhan} compresses  dense document vectors with a jointly trained query encoder and PQ index.
}
Instead of using a sparse inverted index, an alternative retrieval method  is to use a dense representation of each document,
e.g. ~\cite{Lin2021tctcolbert, 2021CIKM-JPQ-Zhan, xiong2021-ANCE,gao-2021-condenser,2021SIGIR-Zhan-ADORE-dense,2021EMNLP-Ren-RocketQAv2}.
We use BM25 because it is a standard reference point. We have also used uniCOIL  for passage re-ranking because a uniCOIL-based sparse retriever is fairly
efficient and its tested relevance result 
is comparable  to that  of  the end-to-end ColBERT as a dense retriever
and other learned sparse representations mentioned above.
Certainly CQ is applicable  for re-ranking with any of dense or sparse retrievers or their hybrid combination.

\textbf{Model numbers cited from other papers.}
As marked in Tables~\ref{tab:overall_psg} and ~\ref{tab:overall_doc}, 
for DeepCT, JPQ and TILDEv2,
we copy the relevance numbers reported in their papers. 
For TCT-ColBERT(v2), DeepImpact and uniCOIL, we obtain their performance using the released checkpoints of 
Pyserini~\footnote{https://github.com/castorini/pyserini/}.  For PreTTR~\cite{MacAvaney2020SIGIR-prettr} on the
passage task and BERT-base on the document task, we cite the relevance performance reported in ~\citet{Hofsttter2020marginMSE}. There are two reasons to list  the relevance numbers from other papers.  One reason is that
for some chosen algorithms,  the running of our implementation version or their code   delivers a performance lower than what has been reported in the authors' original papers, 
perhaps due to the difference in training setup. Thus, we think it is fairer to report the results from   
the  authors' papers. Another reason is that for 
some algorithms, the authors did not release code and we do not have implementations.

In storage space estimation of Table~\ref{tab:storage}, for BECR, we use the default 128 bit LSH footprint with 5 layers. For PreTTR we uses 3 layers with dimension 768 and 
two bytes per number following ~\citet{Hofsttter2020marginMSE}. For TILDEv2, we directly cite the space cost from its paper. 

\textbf{Model implementation and training.}
For baseline model parameters, we use the recommended set of parameters from the authors' original papers. 
For ColBERT, we use  the default version that the authors selected for fair comparison. 
The ColBERT code follows  the original 
version released~\footnote{https://github.com/stanford-futuredata/ColBERT} and BERT implementation is from 
Huggingface~\footnote{https://huggingface.co/transformers/model\_doc/bert.html}.
For BERT-base and ColBERT, training uses  pairwise softmax cross-entropy loss over the released or derived 
triples in a form of ($\textbf{q}, \textbf{d}^+, \textbf{d}^-)$ for the MS MARCO passage task. 
For the MS MARCO document re-ranking task, we split each positive long document  into segments with 400 tokens each and transfer
the positive  label of such a document to each divided segment.
The negative samples are obtained using the BM25 top 100 negative documents. 
The above way we select training triples for document re-ranking may be less ideal and can deserve an improvement in the future.

When training ColBERT, we use gradient accumulation and perform batch propagation every 32 training triplets. 
All models are trained using Adam optimizer~\citep{Kingma2015Adam}. The learning rate is 3e-6 for ColBERT and 2e-5 for 
BERT-base following 
the setup in its original paper. For ColBERT on the document dataset, we obtained the model checkpoint from the authors. 

Our CQ implementation leverages the open source 
code~\footnote{github.com/mingu600/compositional\_code\_learning.git} for~\citet{Shu2018ICLR}.
For PQ, OPQ, RQ, and LSQ, we uses off-the-shelf implementation from 
Facebook's faiss\footnote{https://github.com/facebookresearch/faiss} 
library~\citep{JDH17faiss}.  
To get training instances for each quantizer, we generate 
the contextual embeddings of randomly-selected 500,000  tokens from passages or  documents using ColBERT. 

When using the MSE loss, learning rate is 0.0001, batch size is 128, and the number of training epochs is 200,000. 
When fine-tuning with PairwiseCE or MarginMSE, we
freeze the encoder based  on the MSE loss, set the learning rate to be 3e-6,  and then
train for additional 800 batch iterations with 32 training pairs per batch. 


\end{document}